# Anapole-Mediated Emission Enhancement in Gallium Nitride Nanocavities

*Hao Wang,*[†,‡] *Jing Wang,*[‡] *Shasha Li,*[†,‡] *Kwai Hei Li,*[§] *Hai-Qing Lin,*[‡] *Lei Shao*[†,‡,∥,]*

[†]Shenzhen JL Computational Science and Applied Research Institute, Shenzhen 518131, China.

[‡]Beijing Computational Science Research Center, Beijing 100193, China.

[§]School of Microelectronics, Southern University of Science and Technology, Shenzhen 518055, China.

[∥]State Key Laboratory of Optoelectronic Materials and Technologies, Sun Yat-sen University, Guangzhou 510275, China.

Corresponding author: shaolei@csrc.ac.cn

**ABSTRACT:** Benefiting from their low-loss light manipulation at subwavelength scales, optically resonant dielectric nanostructures have emerged as one of the most promising nanophotonic building blocks. Here, we theoretically conceive a dielectric nanocavity made of moderate-refractive-index gallium nitride and investigate the strong electromagnetic field confinement inside the nanocavity. We demonstrate that gallium nitride nanodisks can support anapole states, which result from interference between electric dipole and toroidal dipole modes and are tunable by changing sizes of the nanodisks. The highly confined electromagnetic field of the anapole states can promote the emission efficiency of a single quantum emitter inside the nanocavity. Moreover, the emission polarization can be tuned by placing the quantum emitter off the nanodisk center. Our findings provide a promising candidate for the construction of ultra-compact, super-radiative integrated quantum light sources.

Single-photon emitters, a type of nonclassical light sources that emit photons one by one and thus exhibit quantum mechanical characteristics, are a fundamental component of numerous quantum photonic applications, such as quantum cryptography, quantum communication, and optical quantum computing.[1–5] To date, a variety of solid-state quantum emitters have been employed to generate single-photon emission, including color centers in crystals, defects in two-dimensional materials, carbon nanotubes, and semiconductor quantum dots.[6–9] However, the extremely small size of a single quantum emitter makes it suffer from intrinsic low emission rate because of the weak interaction with the excitation light. Besides, the photon extraction efficiency of quantum emitters embedded in host materials is usually limited by the small light emitting angle. To overcome such limitations, the single quantum emitter is usually placed in a photonic cavity.[1,10] The cavity allows the confinement of the excitation light energy and thus improves the quantum emitter excitation efficiency. The cavity mode also modifies the optical density of states around the quantum emitter, thus significantly enhancing the spontaneous emission rate, known as Purcell effect.[11] Moreover, the optical cavity can ameliorate the outcoupling efficiency and directionality of photon emission from the sources and as a result increase the extraction efficiency of the embedded emitter. In order to achieve a large Purcell factor, optical cavities with high quality factors as well as small mode volumes are strongly desirable. Several types of optical microcavities have been exploited as high-$Q$ cavities, such as micropillar microcavities with distributed Bragg reflectors (DBRs), microdisk cavities, and photonic crystal cavities.[5,12–15] Because of the narrow linewidth of the optical mode in these optical microcavities, the application of these structures usually requires a precise spectral matching. Unfortunately, the relatively large structure volumes of the microcavities prevent their use in constructing ultracompact on-chip integrated photonic sources. Recently, much attention has been paid to plasmonic metal nanostructure cavities,[16–18] which have ultrasmall mode volumes but always suffer from large metal Ohmic losses.[19]

In the past decade, dielectric nanostructures supporting Mie resonances have emerged as one promising alternative to complement the plasmonic structures.[20–31] Compared with their plasmonic counterparts, dielectric nanocavities can simultaneously exhibit strong electric and magnetic multipolar resonances with much lower optical losses, allowing the enhancement of light emission of both electric and magnetic dipole quantum emitters.[32–34] Furthermore, strong electromagnetic field is not only localized on the surface of dielectric nanocavities like their plasmonic counterparts, but also confined to the interior of the nanocavities. Therefore, one can place quantum emitters inside the dielectric nanocavities, protecting them from the detrimental environment.[35,36] For instance, a recent study demonstrated experimentally the emission enhancement of nitride-vacancy centers embedded in diamond nanoparticles by Mie resonances, which shows great potential in the application of quantum light sources.[37] In addition to radiative Mie resonances, nonradiating anapole state generated by the interference between an electric dipole mode and a toroidal dipole mode has also recently been demonstrated in dielectric nanocavities.[38–40] The optical anapole can produce tightly confined electromagnetic field inside the nanocavities, which has been used in various applications, such as nonlinear harmonic generation, nonradiating laser source, enhanced Raman spectroscopy, and nonlinear photothermal effect.[41–46] Although the anapole state is radiationless, its near-field characteristic is beneficial for the enhancement of emission of a single quantum emitter, which however has rarely been realized.[47,48]

Previous studies on dielectric cavities have utilized various semiconductor materials with different refractive indexes, such as silicon, germanium, gallium arsenide, diamond, titanium dioxide, cuprous oxide, and so on.[20–23,37,41,42,49,50] However, a class of semiconductors widely used in our daily life, namely III-nitride semiconductors, have rarely been explored. III-nitride semiconductors are one of the most successful light-emitting materials for solid-state light sources.[51] Recently, III-nitride quantum dots have attracted intensive interest in nonclassical solid-state single-photon sources because of their short emission wavelength and stable room-temperature operation.[52,53] III-nitride semiconductors also possess moderate refractive indexes, which makes such materials promising for the construction of dielectric nanocavities.[54] In addition, wide-band gallium nitride (GaN) exhibiting low optical losses in the short wavelength range is a natural host material for III-nitride quantum dots. These attractive features make III-nitride semiconductors a promising platform for the study of nanocavity-emitter interaction, which has yet to be explored.

In this study, we demonstrate that a GaN nanodisk can work as an efficient optical nanocavity that enhances the light emission of a single quantum emitter. We show that an anapole-like state can be produced in the GaN nanodisk, where the internal electric energy of the nanocavities is maximized while light scattering to far-field is reduced. By tuning the size of the nanodisk, the optical anapole state can be tailored across the visible range. As a proof of concept, we demonstrate the anapole-mediated emission enhancement of a single electric dipole source inside the GaN nanodisk. We observe a one-order enhancement of emission efficiency of the single dipole optical source embedded in the GaN nanocavity, which results from the increase of both Purcell factor and extraction efficiency, in comparison with the same dipole source embedded in GaN thin film. Moreover, despite the symmetric geometry of the GaN disk, the dumbbell-like electric field distribution at the anapole state leads to polarized light emission of the dipole source placed off the disk center.

RESULTS AND DISCUSSION

Nanodisks offer high degree of freedom to tailor multipole optical resonances and therefore are favorable when designing nanophotonic cavities. Several pioneering works have demonstrated that anapole states can be easily observed in high-aspect-ratio dielectric nanodisks under the excitation of plane waves.[38,41–48] From a practical application point of view, lithographic fabrication of the disk structures is easy to implement. We therefore considered a GaN nanodisk and performed finite-difference time-domain (FDTD) calculations to investigate its optical resonant properties (see Method for details of the numerical simulations). We first set the diameter and height of the GaN nanodisk at 400 nm and 50 nm, respectively. The nanodisk suspended in vacuum was excited by a plane wave from top with linear polarization along $x$-axis (Figure 1a). The total scattering efficiency (defined as total scattering cross section divided by the disk's geometrical cross section) spectrum of the nanodisk, as well as its corresponding normalized internal electric energy $W_v$, i.e. electric energy inside the nanodisk, were therefore calculated (Figure 1b). $W_v$ is defined as $W_V = n^2 \int \frac{E^2 dV}{V E_0^2}$, where $n$ is the refractive index of GaN, $E$ is the electric field, $V$ is the volume of the nanodisk, and $E_0$ is the electric field of incident light. A prominent peak of $W_v$ emerges at $\lambda = 490$ nm. Meanwhile, a dip can be observed in the scattering spectrum at around the peak wavelength of $W_v$. Such near-field and far-field spectral features indicate the appearance of a nonradiating state in this GaN nanodisk.

To reveal the origin of the nonradiating state, we decomposed the nanodisk scattering efficiency spectrum into contributions from different multipolar resonance modes by employing the Cartesian multipole decomposition method based on the electromagnetic fields obtained by the FDTD simulation (see Methods for details). We found that the scattering of the GaN nanodisk is mainly contributed from the electric dipole (ED) mode and the magnetic quadrupole (MQ) mode, while the scattering contributions from the magnetic dipole (MD) mode and the electric quadrupole (EQ) mode are significantly smaller (Figure 1c). The sum of the scattering contributions from the above four multipolar modes agrees well with the calculated total scattering spectrum. The ED mode exhibits a minimal scattering right at the wavelength of $W_v$ maximum. Further multipole decomposition indicates that the ED mode contains the contributions from both the Cartesian electric dipole **P** and the toroidal dipole **T**. These two dipole modes exhibit comparable amplitudes and a phase difference of $\pi$ (Figure 1d and e). The destructive interference of the two dipole modes therefore leads to reduced far-field scattering as well as enhanced near-filed confinement, forming a scattering dark state termed as anapole state. Meanwhile, a pronounced MQ mode appears near the anapole state and dominates the total scattering spectrum near the anapole state wavelength. The far-field radiation therefore does not vanish completely, and the scattering dip is not well matched with the anapole state wavelength, i.e. the $W_v$ peak $\lambda = 490$ nm. We also calculated the electromagnetic field distributions at the anapole wavelength (Figure 1f), which resemble the typical anapole patterns reported in previous works.[38] The slight discrepancy between the near-field distributions in the GaN nanodisk and those in high-refractive index dielectric nanodisks can be ascribed to the excitation of the MQ mode in the moderate-index GaN nanodisk.[55] Although the MQ mode contributes to the far-field scattering, it does not affect much the confinement of internal electric energy at the anapole state, because the MQ mode is a sub-radiative resonance and the MQ resonance frequency is away from the anapole state wavelength.

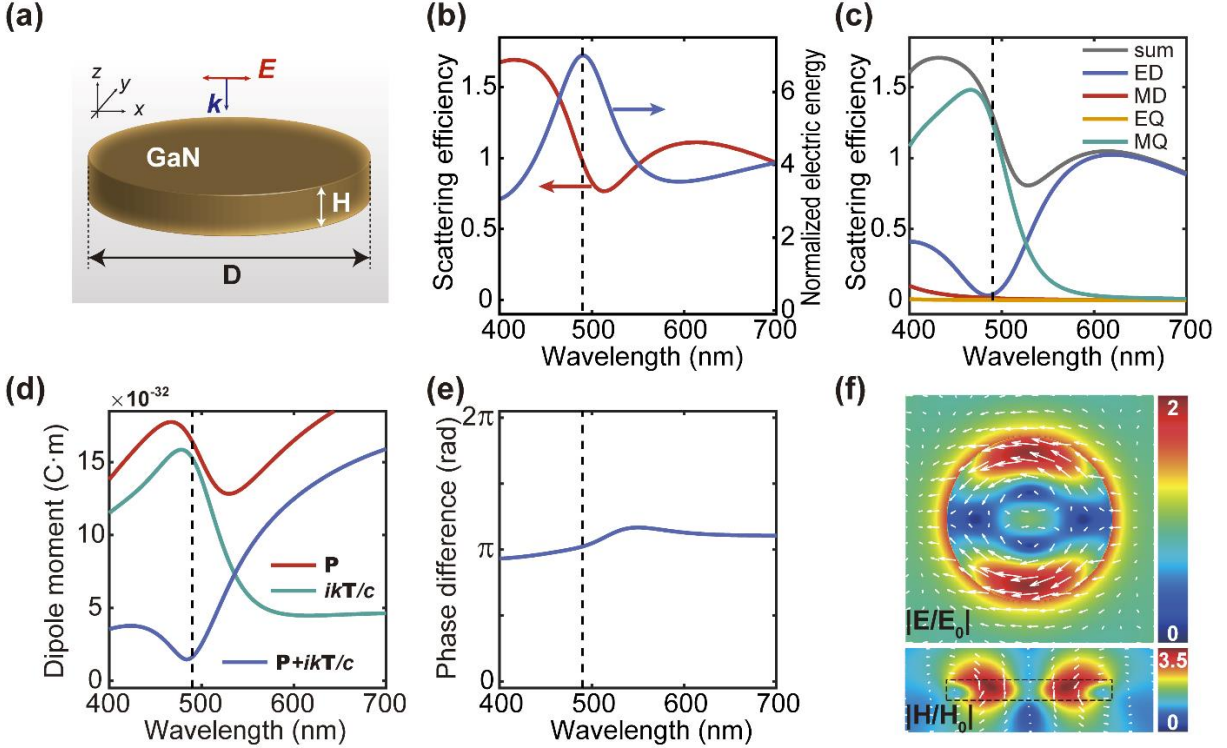

**Figure 1.** Optical response of a GaN nanodisk. The diameter and height of the nanodisk is 400 nm and 50 nm, respectively. (a) Schematic of the GaN nanodisk under the excitation of a linearly polarized plane wave from top. (b) Total scattering efficiency spectrum of the GaN nanodisk (red line) and the spectrum of the corresponding normalized electric energy inside the nanodisk (blue line). The peak position of the internal electric energy spectrum is indicated by a vertical dashed black line. (c) Multipole decomposition of the scattering spectrum of the nanodisk. (d) Dipole moments of the Cartesian electric dipole **P** and toroidal dipole **T** of the GaN nanodisk. (e) Phase difference between the Cartesian electric dipole and toroidal dipole. (f) Electric field enhancement distribution on the central cross section in *xy* plane (upper) and magnetic field intensity enhancement distribution on the central cross section in *xz* plane (lower) at $\lambda = 490$ nm, which corresponds to the peak wavelength of the internal electric energy spectrum. The white arrows show the direction of the electric or magnetic field.

We further calculated the scattering of GaN nanodisks with varying diameters but fixed heights. At a fixed disk height, the scattering dip redshifts linearly with the disk diameter (Figure 2a–c). Besides, the scattering dip also exhibits a red shift when the height is enlarged. We further calculated the spectra of the internal electric energy $W_v$ for the nanodisks. Although the $W_v$ peak does not coincide with the scattering dip, it follows a similar redshift trend as the disk size increases.

To facilitate the investigation of how the nanodisk anapole state modulates the emission properties of an embedded emitter, we carefully adjusted the disk size to ensure that the maximum of $W_v$ appears at 450 nm, which corresponds to the emission peak of commonly used blue III-nitride emitters. The optimized diameters are 346 nm, 298 nm, and 268 nm for the nanodisks with height $H = 50$ nm, 75 nm, and 100 nm, respectively (Figure 2d). As the disk height increases, the diameter-to-height aspect ratio of the optimized disk geometry decreases dramatically, while the disk volume slightly increases. The increasing volume leads to better confinement of the electromagnetic field inside the disk cavity, as reflected by the increasing value of the $W_v$ maximum (Figure 2d), as well as the electric field distribution outside and inside the nanodisk cavity (Supporting Information, Figure S1). The anapole characteristics of the near-field profile gradually disappear as the disk volume increases, because in addition to the MQ resonance, other multipolar resonance modes such as the MD and EQ modes gradually appear at the anapole state wavelength, as suggested by the multipole decomposition results (Supporting Information, Figure S2).

The anapole state is observed generally through the emergence of the far-field scattering dip. However, the scattering dip does not coincide with the peak wavelength of $W_v$ for the GaN nanodisks because of the excitation of other resonance modes (Figure 1b and Figure 2d,e). The difference between the scattering dip and the $W_v$ peak wavelengths depends on the nanodisk size (Figure 2f). Such discrepancy makes it difficult to ascertain the peak wavelength of $W_v$ from optical measurements and therefore brings challenges in designing the optimized structure sizes. Fortunately, we found that the diameter ($D$) and height ($H$) of the GaN nanodisks with $W_v$ peak fixed at a certain wavelength (450 nm in our simulation) follow a power function (Figure 2g),

$$D = aH^b, \qquad (1)$$

where the parameters $a$ and $b$ were obtained by fitting to be 1463 and $-0.3685$, respectively.

Differently sized GaN nanodisks with their internal electric energy maximum fixed near a certain wavelength (450 nm) thus can be designed conveniently. As a proof, we calculated the scattering spectra and corresponding normalized internal electric energy of GaN nanodisks with heights varying from 55 nm to 90 nm and diameters obtained from the above power function (Supporting Information, Figure S3). The calculated $W_v$ peaks of the differently sized nanodisk all fall around the wavelength of 450 nm, as predicted by the design.

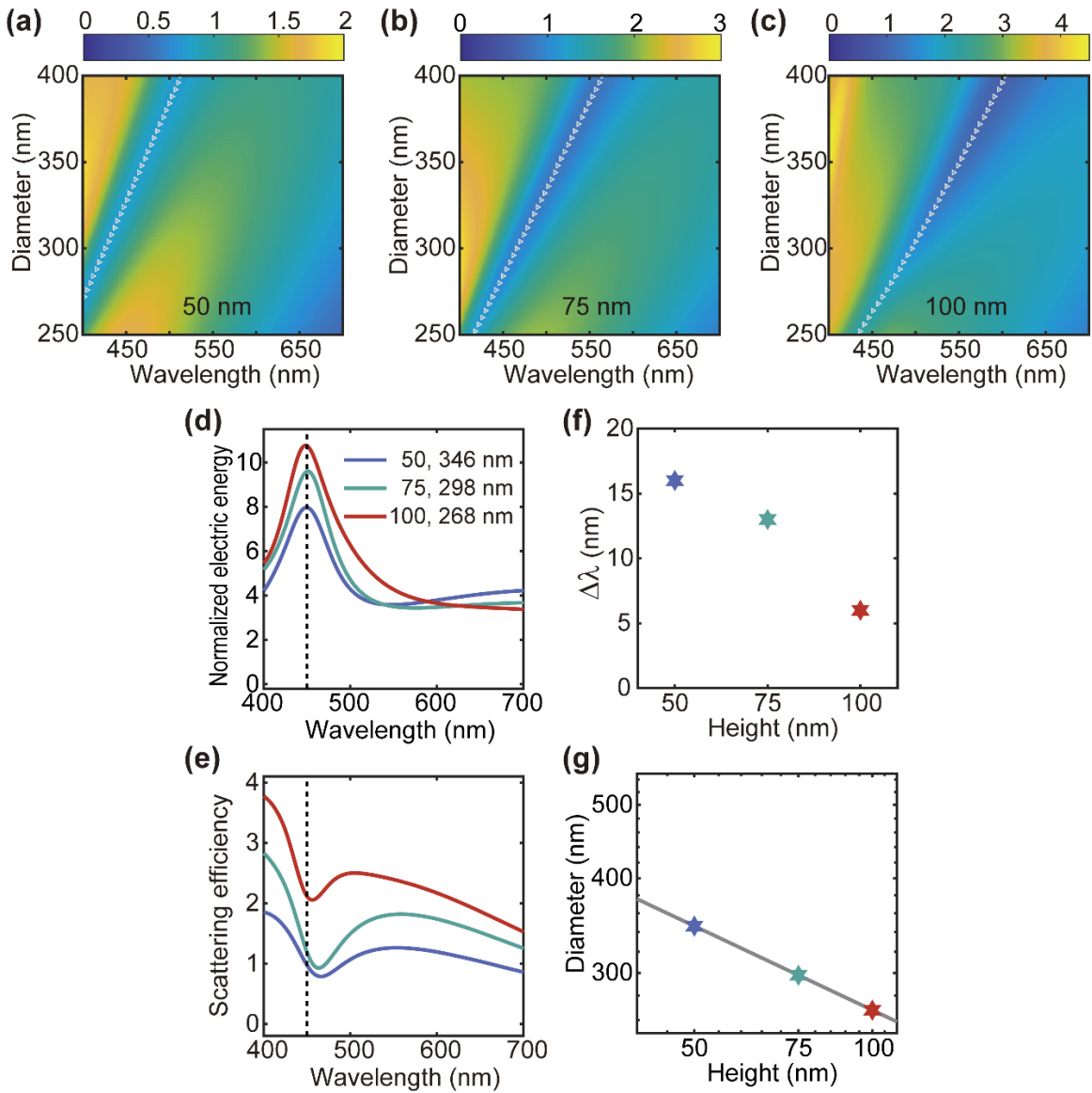

**Figure 2.** Dependence of the anapole state on the size of the GaN nanodisk. (a–c) Scattering

efficiency spectra of the GaN nanodisks with heights of 50 nm (a), 75 nm (b), and 100 nm (c), respectively. The diameters range from 250 to 400 nm. The white marks indicate the scattering dip. (d) Normalized internal electric energy and (e) corresponding scattering efficiency spectra of three typical GaN nanodisks. The diameters/heights of these nanodisks are 50/346 nm, 75/298 nm, and 100/268 nm, respectively. The three nanodisks exhibit the same internal electric energy peak wavelength at 450 nm, which is indicated by a vertical dashed line. (f) Dependence of the wavelength differences between internal electric energy maximum and scattering dip on the heights of the nanodisks. (g) The relationship between the heights and diameters.

The above studies showed the far-field and near-field properties of the anapole state in GaN nanodisks. To investigate how the anapole state modulates the light emission of an embedded single quantum emitter, we further performed FDTD simulations by placing an electric dipole source modelling the single quantum emitter inside the GaN nanodisk cavity. The localized electromagnetic environment in the optical cavity leads to the increase of the spontaneous emission rate of the emitter. The enhancement factor of the decay rate $\Gamma$ in the presence of the cavity in comparison with the decay rate $\Gamma_0$ in the free space, known as Purcell factor, can be expressed as[56]

$$F_\mathrm{P} = \frac{\Gamma}{\Gamma_0} = \frac{P_\mathrm{rad}}{P_0} + \frac{P_\mathrm{nonrad}}{P_0} = F_\mathrm{rad} + F_\mathrm{nonrad} \qquad (2)$$

where $P_\mathrm{rad}$ is the power radiated to far-field, $P_\mathrm{nonrad}$ is the power dissipated in the cavity, and $P_0$ is the emission power of the emitter in free space. Since GaN is lossless at the optical frequencies, the nonradiative decay channel can be neglected, and the Purcell factor can be described by the radiative decay rate enhancement $F_\mathrm{rad}$. In the FDTD simulations, the radiative emission power $P_\mathrm{rad}$ and $P_0$ can be calculated by integrating the Poynting-vector over a closed cubic surface surrounding the nanodisk and the dipole source, respectively.[57]

We chose the GaN nanodisk with $H = 50$ nm and $D = 346$ nm as an example. A dipole source was placed in the nanodisk center (Figure 3a). The polarization of the dipole was set along the *x*- axis, parallel with the localized electric field at the anapole state wavelength. A peak of ~2 is observed at the anapole state wavelength in the spectrum of the radiative decay rate enhancement (Figure 3b), which unambiguously suggests that the emission enhancement of an individual emitter embedded in a dielectric nanocavity is governed by the cavity anapole state.[47,48] The anapole-assisted emission enhancement can be further tuned by varying the size of the nanodisk cavity (Supporting Information, Figure S4). This result seems counterintuitive at first glance since the anapole state is nonradiative. The emission rate of a dipole emitter is determined by the localized density of photonic states (LDOS) modified by the surrounding electromagnetic environment. It is possible to have an enhanced emission if one places only one emission dipole source at a certain point, e.g. the disk center, instead of distributing several dipole sources uniformly in the nanodisk cavity. The nonradiating state dominates in the presence of the interference between different dipole sources. Indeed, we observed a minimized total emission at the anapole state wavelength when we perform calculations to study the total emission of various dipole sources distributed along the $y$-axis of the nanodisk (Supporting Information, Figure S5).

The radiative decay rate enhancement is strongly determined by the position of the dipole source because of the inhomogeneous distribution of the electric field inside the nanodisk when the anapole state is excited (Figure 3a). The enhancement factor at around the anapole wavelength decreases when the $x$-polarized dipole source moves away from the center along the $x$-axis (Figure 3c), which can be understood by the reduced electric field intensity and the declination of electric field direction at the anapole state. The emission spectra of the $x$-polarized dipole source placed on the $y$-axis have different spectral shapes (Figure 3d). When the off-center displacement of the dipole source increases, the enhancement factor decreases first, but increases as the dipole approaches the boundary of the nanodisk. The evolution of the emission rate is in good agreement with the distribution of the electric field. We further compared the radiative decay rate enhancement along the two axes by plotting the enhancement factors at the wavelength of 450 nm as a function of the off-center displacement (Figure 3e). At a displacement smaller than 100 nm, the enhancement factor of the *x*-polarized dipole source placed on the *x*-axis is greater than that of the dipole source placed on the *y*-axis, while at a displacement larger than 100 nm the enhancement factor of the dipole source placed on the *y*-axis is larger.

The two dipole sources located on the *x* and *y* axes with the same displacement from the disk center can be regarded as a dipole source with two mutually perpendicular polarization directions. Therefore, the site-dependent radiative decay rate enhancement indicates a polarized emission characteristic. To further reveal the polarization dependence of the anapole-mediated emission, we calculated the enhancement factors of the dipole source on the *x* axis at $\lambda = 450$ nm for different polarization directions $\theta$ (angle between the dipole source polarization and the *x*-axis). The calculation results showed that the emission enhancement–$\theta$ dependence varies with the off-center displacement. For dipole source placed at the disk center ($x = 0$), the enhancement factor is a constant that does not change with $\theta$. In contrast, when the dipole source is moved away from the center, the enhancement factor becomes $\theta$-dependent (Figure 3f). To quantify the mediation roles of the source polarization on the emission enhancements, we defined the degree of polarization (DOP) as $\rho = (F_{\max} - F_{\min})/(F_{\max} + F_{\min})$, where $F_{\max}$ and $F_{\min}$ are the maximum and minimum enhancement factors, which are achieved at $\theta = 0°$ or $90°$ (Figure 3f). DOP varies with the off-center displacement. As the off-center displacement increases from 0 to 160 nm, the DOP first increases, then deceases, and then increases again. When the dipole source is close to the disk boundary, a large DOP of ~0.75 can be obtained, indicating that an approximate linear polarization state of the emission can be achieved if an unpolarized photon source is placed at the disk boundary.

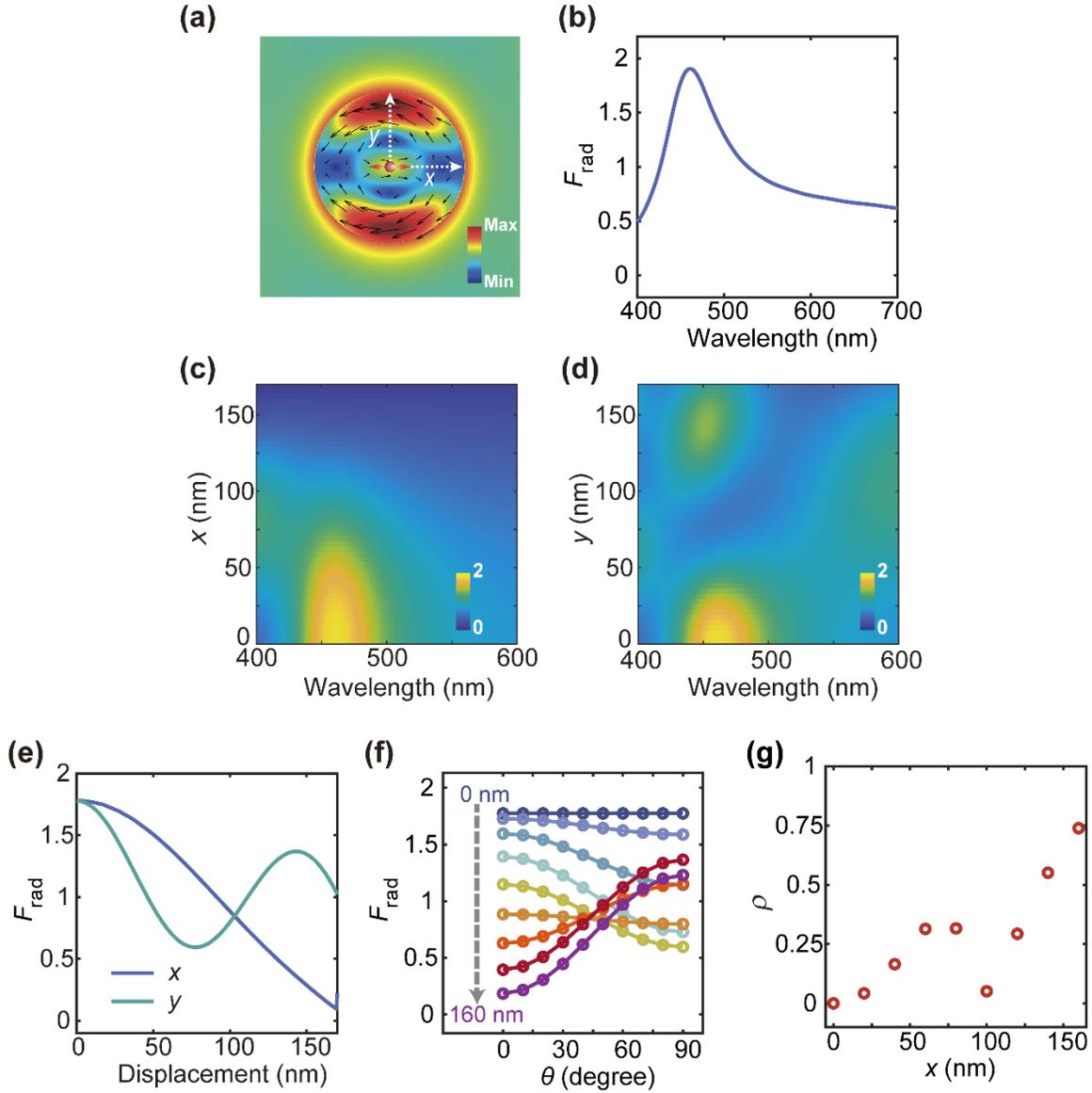

**Figure 3.** Radiative decay rate enhancement in a GaN nanodisk. The height and diameter of the nanodisk are 50 nm and 364 nm, respectively. (a) Electric field intensity enhancement distribution on the central cross section in *xy* plane at the wavelength $\lambda = 450$ nm, where the internal electric energy of the disk cavity reaches its maximum. The nanodisk is excited by a linearly polarized light with its *E* field along the *x* axis. The black arrows indicate the direction of electric field. The schematic of a dipole source placed in the disk center is also plotted. (b) Spectrum of the radiative decay rate enhancement of the dipole source placed in the center of the nanodisk. (c,d) Spectra of the radiative decay rate enhancement of the *x*-polarized dipole source located on the *x*-axis (c) and *y*-axis (d) with different displacements from the disk center. (e) Dependence of the radiative decay rate enhancement at $\lambda = 450$ nm on the off-center displacement when the dipole source is placed on the *x* and *y* axes, respectively. (f) Dependence the radiative decay rate enhancement at $\lambda = 450$ nm on the polarization direction, $\theta$, of the dipole source. $\theta$ is defined as the angle between the polarization direction of the dipole source and the *x*-axis. The dipole source is placed on the *x* axis with the off-center displacement ranging from 0 to 160 nm. (g) Degree of polarization as a function

of the off-center displacement.

It is known that the emission angle of emitters in high-index host materials is very small because of the internal reflection, which greatly limits photon extraction efficiencies of quantum emitters. Optical nanocavities can not only increase the Purcell factor but also improve the outcoupling efficiency of the emitters. Therefore, despite the moderate Purcell factor, the total enhancement of the observable light emission is still considerable when an optical nanocavity is employed. It should be noted that in experiments the optical nanocavities are always placed on a substrate. The commonly used substrate for GaN is sapphire. However, because of the relatively small refractive-index contrast between GaN and sapphire, the GaN nanostructures supported on sapphire substrates cannot achieve a strong light confinement anymore. The anapole state therefore cannot be supported (Supporting Information, Figure S6). To avoid such scenario, we designed a suspended structure by placing the nanodisk on a cone-shaped sapphire nanopillar (the inset of Figure 4).[58,59] The suspension of the GaN nanodisk redistributes the electromagnetic field and prevents the GaN nanodisk cavity from severe leakage of the confined light energy. In our simulation, the height of the cone-shaped sapphire nanopillar is 100 nm, and the top and bottom diameters are 50 nm and 100 nm, respectively. The size of the GaN nanodisk is the same as that in Figure 3. Such suspended GaN nanocavity maintains the anapole state (Supporting Information, Figure S6), which contributes to the enhanced emissions. We defined a total emission enhancement factor, $F_{total}$, as the ratio between the emission power of a light source inside the suspended GaN nanodisk and the emission power of the same light source in a corresponding GaN film. The emission power is calculated by integrating the Poynting-vector over a surface on top of the structures. Our calculation showed that the emission power of a dipole source can be enhanced by an order of magnitude when the dipole source is placed in the center of the suspended GaN nanocavity, with $F_{total}$ reaches around 15 (Figure 4). Our results clearly verified the effectiveness of the nanocavity suspension strategy, in which the light emission of single quantum emitters is enhanced by both improved Purcell factor and photon extraction efficiency.

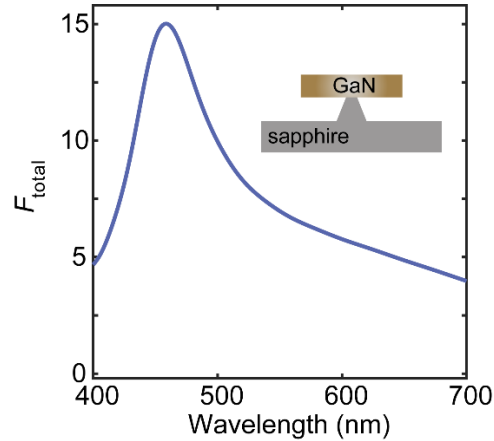

**Figure 4.** Total emission enhancement of a dipole source placed in the center of the suspended GaN nanodisk cavity. Inset shows the schematic of the suspended nanodisk.

CONCLUSION

In conclusion, we have theoretically investigated the excitation of anapole resonance of moderate-index GaN nanodisks and demonstrated the enhanced light emission of single quantum emitters by the anapole states of the nanodisk cavity. The nonradiating anapole state produces strong electromagnetic field enhancement inside the nanodisk cavity, leading to ~2-fold Purcell factor enhancement of a single quantum emitter placed in the cavity center. Additionally, the nanodisks can serve as compact nanoantennas to improve the light extraction efficiency of quantum emitters. As a result, one order of magnitude enhancement in emission efficiency can be achieved by employing the GaN nanodisk cavity. Moreover, the dipole emitter inside the GaN nanodisk cavity exhibits polarization-dependent emission with a degree of polarization reaching 0.75 when the emitter is placed off the cavity center. We further proposed the experimental realization of the GaN anapole resonator by employing the suspended nanodisk cavity and analyzed the emission enhancement performance. Our design can be extended to nanocavities made of other moderate-refractive-index and low-refractive-index dielectric materials. We believe our work affords a promising path towards the realization of compact and ultra-radiative integrated quantum light sources.[47,48]

METHODS

**FDTD Simulations.** The numerical simulations were performed using a commercial FDTD software (Lumerical FDTD Solutions). The dielectric function of GaN was obtained from a previous work.[54] The surrounding environment was set as vacuum with refractive index of 1. The refractive index of the sapphire substrate was 1.78. Perfectly matched layers were used at the boundary to absorb the scattered radiation. To calculate the scattering spectra and near-field distributions, the structures were excited by linearly polarized total-field scattered-field plane wave. To calculate the radiative decay rate, an electric dipole was used as the excitation source. The mesh size was set to 2.5 nm around the nanodisk.

**Cartesian Multipole Decomposition.** The Cartesian multipole moments of a nanoparticle are determined by the current density $\mathbf{J} = -i\omega\varepsilon_0(\varepsilon_r - 1)\mathbf{E}$ induced in the nanostructure, where $\omega$, $\varepsilon_0$, $\varepsilon_r$ and $\mathbf{E}$ are the angular frequency, vacuum permittivity, relative permittivity of the material, and electric field, respectively. The field data were obtained from the FDTD simulations. The electric dipole moment $\mathbf{P}$, toroidal dipole moment $\mathbf{T}$, magnetic dipole moment $\mathbf{M}$, electric quadrupole moment $\mathbf{Q^E}$, and magnetic quadrupole moment $\mathbf{Q^M}$, can be calculated by

$$\mathbf{P}=\frac{i}{\omega}\int \mathbf{J}d^3r$$

$$\mathbf{T} = \frac{1}{10}\int [(\mathbf{r}\cdot \mathbf{J})\mathbf{r} - 2r^2\mathbf{J}]d^3r$$

$$\mathbf{M} = \frac{1}{2}\int (\mathbf{r}\times \mathbf{J})d^3r$$

$$\mathbf{Q}^{\mathrm{E}}_{\alpha,\beta} = \frac{i}{\omega}\int (r_\alpha J_\beta + r_\beta J_\alpha - \frac{2}{3}\delta_{\alpha,\beta}(\mathbf{r}\cdot \mathbf{J}))d^3r$$

$$\mathbf{Q}^{\mathrm{M}}_{\alpha,\beta} = \frac{1}{3}\int ((\mathbf{r}\times \mathbf{J})_\alpha r_\beta + (\mathbf{r}\times \mathbf{J})_\beta r_\alpha)d^3r,$$

where **r** is the displacement vector from the origin at the center of the nanostructure to the point of current distribution area. The scattering power can be written as

$$P_{\mathrm{scat}} = P_{\mathrm{scatED}} + P_{\mathrm{scatMD}} + P_{\mathrm{scatEQ}} + P_{\mathrm{scatMQ}}$$

$$= \frac{k^4}{12\pi\varepsilon_0^2 c\mu_0}\left|\mathbf{P}+\frac{ik}{c}\mathbf{T}\right|^2 + \frac{k^4}{12\pi\varepsilon_0 c}|\mathbf{M}|^2 + \frac{k^6}{1440\pi\varepsilon_0^2 c\mu_0}\sum_{\alpha,\beta}|\mathbf{Q}^{\mathrm{E}}_{\alpha,\beta}|^2$$

$$+ \frac{k^6}{160\pi\varepsilon_0 c}\sum_{\alpha,\beta}|\mathbf{Q}^{\mathrm{M}}_{\alpha,\beta}|^2$$

where $k$ is the wavenumber, $\mu_0$ is the vacuum permeability, and $c$ is the light speed in vacuum, respectively. The scattering cross section is obtained by dividing the scattering power by the incident energy flux as

$$\sigma_{\mathrm{scat}} = 2\sqrt{\frac{\mu_0}{\varepsilon_0}}\frac{P_{\mathrm{scat}}}{|\mathbf{E}_0|^2}$$

where $\mathbf{E}_0$ is the electric field of the incident light. The corresponding scattering efficiency is then defined as ratio between the scattering cross section and the geometry cross section of the nanodisk

$$Q_{\mathrm{scat}} = \frac{\sigma_{\mathrm{scat}}}{\sigma_{\mathrm{geom}}}$$

ASSOCIATED CONTENT

Supporting Information

The Supporting Information is available free of charge at https://pubs.acs.org/doi/xxxx.

Figures S1–S6 (PDF).

AUTHOR INFORMATION

The authors declare no competing financial interest.

ACKNOWLEDGEMENTS

This work was financially supported by National Natural Science Foundation of China (Grant No. NSAF U1930402), China National Postdoctoral Program for Innovative Talents (Grant No. BX20200223), China Postdoctoral Science Foundation (Grant No. 2020M682898), and State Key Laboratory of Optoelectronic Materials and Technologies of China (Grant No. OEMT-2019-KF-07). L.S. acknowledges the support from the Pearl River Talent Recruitment Program (2019QN01C216) and computational resources from the Beijing Computational Science Research Center.